\documentclass[conference]{IEEEtran}
\IEEEoverridecommandlockouts
\usepackage{cite}
\usepackage{amsmath,amssymb,amsfonts}
\usepackage{algorithmic}
\usepackage{graphicx}
\usepackage{textcomp}
\usepackage{xcolor}
\def\BibTeX{{\rm B\kern-.05em{\sc i\kern-.025em b}\kern-.08em
    T\kern-.1667em\lower.7ex\hbox{E}\kern-.125emX}}
\begin{document}

\title{AMNet: An Acoustic Model Network for Enhanced Mandarin Speech Synthesis}
\author{
\IEEEauthorblockA{Yubing Cao$^1$, Yinfeng Yu$^1$$^{\dagger}, $Yongming Li$^1$$^{\dagger}$, Liejun Wang$^1$\thanks{$^{\dagger}$Both Yinfeng Yu and Yongming Li are corresponding authors.} }
\\
\IEEEauthorblockA{$^1$Xinjiang Multimodal Intelligent Processing and Information Security Engineering Technology Research Center, \\ School of Computer Science and Technology, Xinjiang University, China}
\\
\IEEEauthorblockA{E-mail: yuyinfeng@xju.edu.cn, Lymxju@xju.edu.cn}
}


\maketitle

\begin{abstract}
This paper presents AMNet, an Acoustic Model Network designed to improve the performance of Mandarin speech synthesis by incorporating phrase structure annotation and local convolution modules. AMNet builds upon the FastSpeech 2 architecture while addressing the challenge of local context modeling, which is crucial for capturing intricate speech features such as pauses, stress, and intonation. By embedding a phrase structure parser into the model and introducing a local convolution module, AMNet enhances the model’s sensitivity to local information. Additionally, AMNet decouples tonal characteristics from phonemes, providing explicit guidance for tone modeling, which improves tone accuracy and pronunciation. Experimental results demonstrate that AMNet outperforms baseline models in subjective and objective evaluations. The proposed model achieves superior Mean Opinion Scores (MOS), lower Mel Cepstral Distortion (MCD), and improved fundamental frequency fitting \( F0 (R^2) \), confirming its ability to generate high-quality, natural, and expressive Mandarin speech. 
\end{abstract}


\section{Introduction}
Speech synthesis converts text to natural and expressive speech and is widely used in applications like virtual assistants, intelligent reading, and smart navigation. Recently, Deep Neural Networks (DNNs)\cite{b1,b2,b3,b4,b5,b6} have significantly advanced speech synthesis, particularly through Sequence-to-Sequence (Seq2Seq)\cite{b7} models with attention mechanisms, improving Text-to-Speech (TTS) performance. For instance, Wang et al.\cite{b8} proposed an end-to-end TTS model based on the Seq2Seq attention mechanism, followed by models like Char2wav\cite{b9}, Tacotron\cite{b10}, and Deep Voice 3\cite{b11}, which further enhanced TTS technology.

Despite the progress made with Auto-Regressive (AR) models, they have limitations in computational efficiency and parallelism due to reliance on Recurrent Neural Networks (RNNs)\cite{b12}. Non-Auto-Regressive (NAR) models, such as FastSpeech\cite{b13} and FastSpeech2\cite{b14}, address these issues, speeding up synthesis and capturing global context. However, NAR models struggle with attention dispersion in the self-attention mechanism, which weakens their ability to focus on local relationships between adjacent words or characters\cite{b15}.

Although Transformer-based methods enhance parallel computation and variable-length sequence processing, their main drawback is poor local modeling capability. The self-attention mechanism captures global context but may overlook local dependencies, which are crucial in Mandarin, where the pronunciation of a character depends on its adjacent characters\cite{b16}. Insufficient modeling of these contextual relationships can lead to less fluent and natural speech.

Tone labels are essential in Mandarin Chinese for ensuring accurate pronunciation and conveying meaning\cite{b17}. As a tonal language, Mandarin's tonal variations are critical for semantic distinction. For example, the syllable "wa" (wā, wá, wǎ, wà) conveys different meanings depending on the tone. Despite advances in acoustic feature-based models, there is limited research on the phonemic structure of Mandarin and its relationship with pronunciation. The deeper connection between tone features and pronunciation remains an area for further exploration.

Building on previous research, this paper proposes AMNet, an acoustic model method for Mandarin speech synthesis based on phrase structure annotation and local convolution, to address the limitations of local modeling capability in the self-attention mechanism. Our approach is based on sentence structure analysis and local context modeling. Specifically, we first analyze the structure of Mandarin sentences and propose a phrase structure parser (PSP) to enrich the linguistic information in the input sequence. These phrase annotations are embedded as auxiliary features into the input sequence to enhance the model's ability to perceive local sentence boundaries. Secondly, we introduce a one-dimensional convolutional network (1D-CNN) into the self-attention mechanism to generate feature maps for the query and value, thereby improving the model's sensitivity to local context information. The model can capture semantic correlations between local contexts by calculating the dot product between the query and key, further optimizing the local modeling effect. Additionally, we decouple Mandarin phonemes by treating vowels' tonal characteristics as independent conditional inputs to the model, allowing explicit modeling of specific tones during training and prediction, effectively improving the accuracy of speech synthesis pronunciation.

The main contributions of this paper are summarized as follows:

\begin{itemize}
\item We designed a local modeling enhancement method based on the self-attention mechanism. Using a phrase structure parser module to analyze sentence structure and enrich linguistic information, these annotations are embedded into the input sequence to enhance local boundaries. Additionally, incorporating local convolution operations significantly improves the model's ability to model local context information.

\item We decouple the tonal characteristics in Mandarin phonemes as independent feature conditions, providing explicit guidance for model training and prediction, thereby enhancing the accuracy of tone synthesis and improving the naturalness of the synthesized speech.

\item Experimental results show that the proposed AMNet model demonstrates excellent performance in speech synthesis, with significant improvements in the naturalness of pauses, stress, and intonation.
\end{itemize}

\section{Related Work}
Tone is critical in speech synthesis, directly influencing tonal expression and semantic clarity. Several studies have focused on extracting and modeling tone features, advancing the development of speech synthesis technology. 
Debnath et al \cite{b18}. proposed a method for extracting Mel-Frequency Cepstral Coefficients (MFCC), mainly focusing on features related to pitch and tone. Liu et al \cite{b19}. introduced a tone learning approach for bilingual low-resource speech synthesis, improving intelligibility and accent accuracy. Ekpenyong et al\cite{b20}. used syllables as the synthesis unit and developed a more compelling tone language modeling system.

Zhu et al\cite{b21}. evaluated TTS models’ ability to learn Mandarin tone co-articulation, showing that the models could generate high-quality speech while extracting linguistic knowledge from coarse-labeled data. Lu et al\cite{b22}. proposed an end-to-end Chinese TTS system based on Tacotron2\cite{b23} and WaveNet\cite{b24}, which addresses issues with large character sets and prosodic phrasing, though it treats tone as part of the phone rather than decoupling it explicitly.

Regarding pronunciation accuracy, ToneNet\cite{b25}, a CNN-based model, uses Mel spectrograms and classifies Mandarin syllables into four tones, achieving high accuracy and F1 scores. Wu et al\cite{b26}. improved tone modeling by using phoneme and tone sequences as input, mapping each phoneme’s tone to its corresponding syllable tone.
\begin{figure*}[ht]
\centering
\includegraphics[width=1\textwidth]{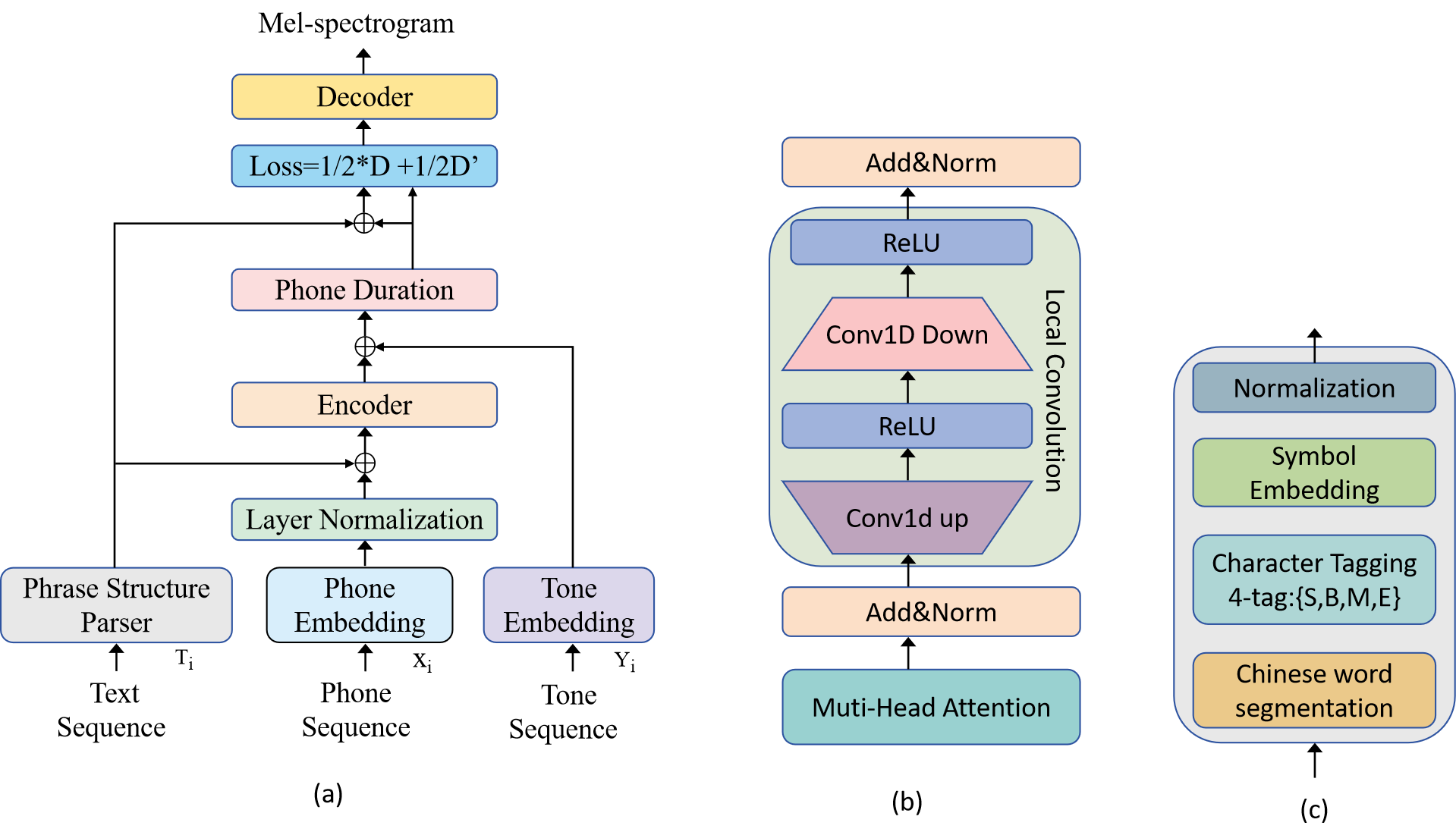}
\caption{(a). AMNet architecture. (b). Encoder architecture. (c). Phrase structure parser.}
\label{1}
\end{figure*}

Phrase annotation is an essential preprocessing step in natural language processing (NLP) tasks\cite{b27}, primarily used to address problems such as text classification and syntactic analysis. Character-based annotation and word embedding-based approaches are common examples among the phrase annotation methods. Character-based annotation, a simple and effective method for Chinese word segmentation, is widely applied in text processing\cite{b28}. It assigns SBME (Single Element, Begin, Middle, End) labels to Chinese characters, explicitly marking their positions within phrases, thereby effectively representing phrase length and positional information\cite{b29}. This method has demonstrated high accuracy in Chinese word segmentation tasks.Reference\cite{b29} introduced the SBME-based annotation method, which combines phrase length and positional information to significantly reduce word tagging errors during the segmentation process, providing an efficient solution for Chinese text processing. However, traditional phrase annotation methods are typically used as a system's frontend text processing module, often overlooking the impact of phrase length and positional information on deep feature extraction. This, to some extent, limits the model's ability to capture contextual semantics.

In addition, in other natural language processing tasks, such as homophone disambiguation in speech synthesis, part-of-speech (POS) tagging also plays an important role. Although the objectives of POS tagging differ from those of phrase annotation, it provides crucial support for the disambiguation of polyphonic characters and semantic analysis in NLP systems. This indicates that phrase annotation and POS tagging each serve different functions, but both contribute, to some extent, by providing semantic and structural information for downstream tasks. The design and optimization of phrase annotators directly assist tasks such as word segmentation and lay a solid foundation for Chinese semantic understanding and feature modeling.
\section{Method}
\subsection{Skeleton structure of the framework}\label{AA}
The AMNet architecture proposed in this paper is based on FastSpeech2, as shown in Fig.~\ref{1}(a), with modifications tailored for Mandarin speech synthesis tasks. The encoder integrates a local convolutional network to enhance the model's ability to capture local context information. Phrase structure is extracted by a phrase structure parser, which is used for phoneme duration correction and encoding. Additionally, to further improve the model's functionality, two new modules are introduced: the phrase structure parser, which captures phrase structure information from the text and integrates phrase-level linguistic features into the model, enriching the linguistic information in the input sequence, and the tone embedding module, which enhances the model's ability to capture and represent tone information, optimized explicitly for the importance of tones in Mandarin, thereby improving the accuracy and naturalness of speech synthesis. During the training phase, a dataset containing text-phoneme-audio triplets is used. Specifically, the training dataset can be represented as ({T, X, M}), where (T) denotes the text sentence, (X) denotes the corresponding phoneme sequence, and (M) represents the speech representation, i.e., the Mel-spectrogram. Furthermore, given a known phoneme sequence (X), we can further decouple it to obtain the corresponding tone sequence (Y), providing explicit tone information to guide the model.
\subsection{Phrase Structure Parser}
\begin{figure*}[ht]
\centering
\includegraphics[width=1\textwidth]{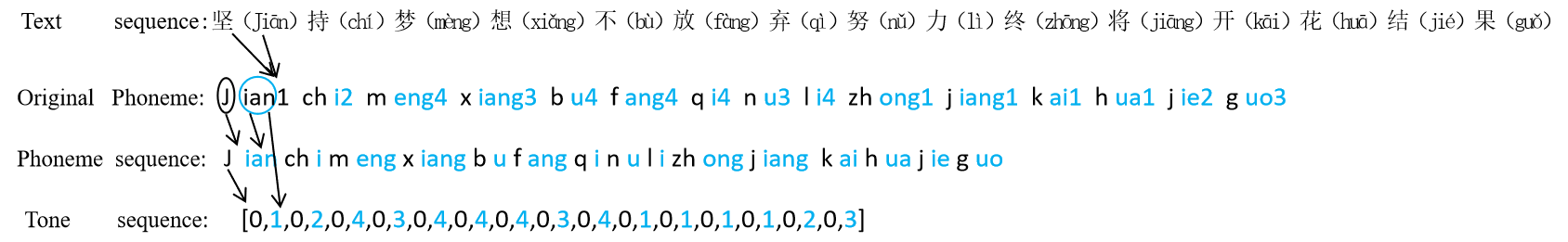}
\caption{The processing of intonation sequences.}
\label{2}
\end{figure*}

The structure of PSP is shown in Fig.~\ref{1}(c). In Mandarin speech synthesis tasks, the input sequence consists of a phoneme sequence \( X = (x_1, x_2, \dots, x_n) \), Chinese text \( T \), and tone sequence \( Y \). As shown in Figure 1(c), the text sequence \( T \) is directly input into the phrase structure parser for phrase structure extraction. The phrase structure extraction task is divided into two steps: first, the text sentence \( T \) is decomposed into phrases using HMM\cite{b30} and the Viterbi algorithm\cite{b31}; then, the phrases are annotated to identify the local boundaries of the phrases.

In the phrase annotation task, different symbols represent the specific positions of Chinese characters within the phrase, and the annotation results are converted into a segmented text sequence. Inspired by Chinese word segmentation (CWS), this paper proposes a character annotation method that uses four labels to construct a phrase boundary annotation set. The four-label set consists of \( \{ S, B, M, E \} \), where \( S \), \( B \), \( M \), and \( E \) represent a single character, the first character of a phrase, a middle character of a phrase, and the last character of a phrase, respectively. Based on the annotation results, the character annotation sequence is further converted into the corresponding numerical annotation sequence. For example: \( \{ B, E, B, E, E, S, \dots \} \Rightarrow (2, 4, 2, 4, 2, 4, 1, \dots) \). Subsequently, symbol embedding and normalization are performed. Each character's label is mapped to the corresponding phoneme, forming the input to the encoder.

As shown in Fig.~\ref{1}(a), under the guidance of the phrase annotator's output, the duration vector of phonemes \( D = [x_1, x_2, \dots, x_{n-1}, x_n] \) is summed to obtain the phrase-level duration vector \( D' = [x_1 + x_2 + \dots, \dots, x_{n-1} + x_n] \). This phrase-level duration serves as an additional constraint during model training, and enhancing duration prediction further improves the model's performance.

\subsection{Local Convolution}
Fig.~\ref{1}(b) shows the local convolution mechanism, where a two-layer one-dimensional convolutional neural network (1D CNN) with ReLU activation function is used to extract features from the text sequence. The convolution kernel size of the one-dimensional convolutional layer is defined as \( h \times d \), where \( h \) and \( d \) represent the height and width of the kernel, respectively. The width \( d \) is typically equal to the dimension of the input phoneme embedding. Let the input matrix be \( A \), and the kernel parameters be \( W \in \mathbb{R}^{d \times h} \), the convolution operation can be expressed as: 
\begin{equation}
o_i = W A[i : i + h - 1]\label{eq1},
\end{equation}
here, \( o_i \) represents the convolution result at the \( i \)-th position. Further introducing the activation function \( f \) (ReLU activation is used in this paper) and the bias \( b \), the final convolution output can be expressed as:
\begin{equation}
c_i = f(o_i + b)\label{eq2}.
\end{equation}

For a 1D CNN with a single convolutional kernel, the size of the resulting feature vector after convolution can be expressed as: 
\begin{equation}
c \in \mathbb{R}^{s-h+1}\label{eq3},
\end{equation}
where \( s \) represents the length of the input sequence, and the number of features is \( s-h+1 \).

After extracting high-dimensional latent representations from the 1D convolutional layer, it is necessary to ensure that the input and output dimensions remain consistent. To achieve this, the same strategy as in \cite{b32} is adopted, where the kernel size \( h \), stride \( s \), and padding \( p \) of the convolutional layer are carefully designed. Let the number of phonemes in the input sequence be \( m \), then the length of the output sequence can be expressed as \( \frac{m + 2p - h}{s} + 1 \).

In this study, the embedding dimension of the input phoneme sequence is 256, i.e., the input matrix \( A \in \mathbb{R}^{h \times 256} \). Three convolutional kernels with different region sizes are used, with sizes \( W_1 \in \mathbb{R}^{9 \times 256} \), \( W_2 \in \mathbb{R}^{5 \times 256} \), and \( W_3 \in \mathbb{R}^{3 \times 256} \). To ensure that the convolved features have the same length as the input, the padding values are set to 4, 2, and 1, respectively. The convolution process expands the embedding dimension from 256 to 1024. Then, three convolutional kernels with region sizes \( W_1' \in \mathbb{R}^{1 \times 256} \), \( W_2' \in \mathbb{R}^{1 \times 256} \), and \( W_3' \in \mathbb{R}^{1 \times 256} \) are applied to compress the features back to 256 dimensions. At this stage, the padding for all convolutional kernels is set to 0. Finally, the features extracted by the three different convolutional kernels are averaged and fused, as shown in the following formula:
\begin{equation}
O=\frac{1}{n}\sum_{i=1}^{n}c_i^2(c_i^1)\label{eq4},
\end{equation}
here, \( c_i^1 \) and \( c_i^2\) represent the results of the \( i \)-th convolution, and \( n \) is the number of convolutional kernels. In this paper, the number of convolutional kernels is set to 3.

\subsection{Tone embedding}
In Mandarin, tonal variations are used to distinguish word meanings, including Tone-1 (High Level), Tone-2 (Rising), Tone-3 (Falling-Rising), and Tone-4 (Falling). Additionally, scholars generally accept that there is a fifth neutral tone in Mandarin. In this paper, we consider this soft tone feature in modeling and design the tone annotation accordingly. Specifically, we use the labels \( \{1, 2, 3, 4, 5\} \) to represent Tone-1, Tone-2, Tone-3, Tone-4, and the neutral tone in vowels. For example, when the corresponding tone is annotated with "a," the marking results are: "\( \bar{a} \)" (Tone-1), "\( \acute{a} \)" (Tone-2), "\( \check{a} \)" (Tone-3), "\( \grave{a} \)" (Tone-4), and "a" (neutral tone). For consonants, "0" is used to denote the initial consonant. According to this annotation rule, tones can be separated from the original phonemes, achieving decoupling of tone and phoneme, as shown in Fig.~\ref{2}. For instance, for the original phoneme "wo3", the phoneme processing steps are as follows: First, the consonant part of the original phoneme is kept unchanged. In contrast, the vowel part is split into characters and digits to generate the tone sequence ([0,3]). Then, the initial consonant is recombined with the vowel part (without the tone) to form a new phoneme sequence, and the tone sequence is composed of the numeric label and "0". Through this approach, this paper achieves independent modeling of the tonal part in the phonemes, which helps improve the model’s performance in tone prediction and phoneme generation, thereby enhancing the accuracy and naturalness of speech synthesis.

\section{Experiments}

\subsection{Data configurations}
The experiments in this study were conducted on the publicly available single-speaker dataset BIAOBEI\footnote{https://www.data-baker.com/data/index/TNtts/}. The BIAOBEI dataset, provided by Biaobei Technology Co., Ltd., consists of 10,000 Mandarin speech sentences, each accompanied by corresponding audio recordings and their text annotations. The total effective speech duration of the dataset is approximately 12 hours, with an audio sampling rate of 22.05 kHz. Feature extraction from the speech uses a frame-shift size of 12.5 ms and a window size of 50 ms to generate spectrograms. The corpus is randomly divided into training, validation, and testing in the experiments. Specifically, the training set contains 9,400 sentences, the validation set contains 500 sentences, and the test set contains 100 sentences. This data split provides a reliable experimental foundation for model training and performance evaluation.

In this experiment, we used a server equipped with 4 Tesla T4 GPUs, each with a memory capacity of 16GB. The server's CPU is an Intel Xeon Gold 5218R, and the total number of training steps was 100,000, with a batch size of 64. The optimizer used is Adam, with parameters set to \( \beta_1 = 0.9 \), \( \beta_2 = 0.98 \), and a weight decay coefficient of \( \lambda = 10^{-9} \). The learning rate schedule was set according to the method in reference \cite{b33}. Additionally, gradient clipping was performed every 30,000 steps to stabilize the training process. To reconstruct speech audio from the predicted Mel spectrograms, we trained a HiFi-GAN-based\cite{b34} vocoder. HiFi-GAN is a high-quality neural vocoder, and its standard implementation was applied to both the proposed model and the baseline model to ensure a fair performance comparison.

The model architecture proposed in this paper is based on FastSpeech 2, and its overall framework retains the non-autoregressive structure design from the original model. The hyperparameters of the encoder, decoder, and variance adaptor are consistent with those of FastSpeech 2. Building upon this foundation, we introduce a local convolution module to enhance the model's ability to capture local features. Specifically, the local convolution module uses two different filter designs: upsampling and downsampling. The size of the upsampling filter is \( 9 \times 1 \), and the size of the downsampling filter is \( 1 \times 1 \), with corresponding channel numbers of 1024 and 256, respectively. This design allows the model to better capture local features in the text sequence while restoring high-quality audio information during decoding.

We selected two baseline models for performance comparison to validate the effectiveness of the model proposed in this paper. Both baseline models use phonemes as input and were trained on the BIAOBEI dataset. FastSpeech 2 is a classic non-autoregressive speech synthesis model, and its architecture and hyperparameter design are publicly available. The model structure in this paper follows the main design of FastSpeech 2, with the addition of a local convolution module to further enhance its ability to model regional features. Sag-Tacotron\cite{b35} is a Gaussian mixture local modeling method based on the self-attention mechanism. This model was used initially for speech synthesis tasks, employing a prior attention mechanism to address local modeling issues. Similar to the approach in this paper, Sag-Tacotron also focuses on local feature modeling and has made some synthesized speech samples publicly available, providing reliability for comparing results.

\subsection{Evaluation metrics}
In the subjective evaluation, the Mean Opinion Score (MOS) rating is given by participants to assess the quality of the generated speech, with a score ranging from 1 (very poor) to 5 (excellent). The evaluation focuses on aspects such as the authenticity of the audio quality, the naturalness of pause placement, and the expressiveness of the voice. The ABX preference test requires participants to select the higher quality audio clip from two speech samples or remain neutral if the quality of both clips is similar. All subjective evaluation experiments were conducted in a quiet studio environment, with participants being native Chinese speakers. Each audio clip was listened to at least three times using professional headphones to obtain an average rating.

In the objective evaluation, Mel Cepstral Distortion (MCD) is used to quantify the difference between the predicted Mel cepstral coefficients and the target Mel cepstral coefficients, with smaller values indicating higher similarity between the synthesized speech and the actual speech. Additionally, the fundamental frequency fitting goodness \( F0 (R^2) \)is used to measure the match between the generated speech and the target speech in terms of fundamental frequency. Higher R² values indicate better prosody control performance\cite{b36}. To further visually compare the performance of different models in generating audio, this paper also presents spectrograms and fundamental frequency curves, highlighting the differences in the frequency domain and fundamental frequency features between the synthesized and target speech.

\subsection{Comparison with existing models}
During the evaluation process, the test audio includes synthesized audio generated by the model proposed in this paper and audio generated by the baseline models. Subjective evaluation is conducted using the ABX preference test, combined with MOS, to assess the naturalness and quality of the speech. Objective evaluation uses MCD and fundamental frequency fitting goodness \( F0 (R^2) \)to quantify the quality and feature accuracy of the generated speech. A detailed analysis is provided below.

\subsubsection{Subjective evaluation}
\begin{table}[t]
\caption{Results of The ABX Preference Tests}
\begin{center}
\begin{tabular}{|c|c|c|c|c|}
\hline
\textbf{} & FastSpeech2 & Sag-Tacotron & \textbf{AMANet(Ours)} & No Preference \\
\hline
1 & 8.7\% & / & \textbf{73.4\%} & 17.9\% \\
\hline
2 & / & 31.2\% & \textbf{49.5\%} & 19.3\% \\
\hline
\end{tabular}
\label{表1}
\end{center}
\end{table}

\begin{table}[t]
\caption{MOS Evaluation Results}
\begin{center}
\begin{tabular}{|c|c|}
\hline{Model} & {MOS($\uparrow$)} \\
\hline
Ground truth & 4.43$\pm$0.12 \\
\hline
FastSpeech2 & 3.72$\pm$0.12 \\
\hline
Sag-Tacotron & 3.97$\pm$0.13 \\
\hline
\textbf{AMNet (Ours)} & \textbf{4.11$\pm$0.12} \\
\hline
\end{tabular}
\label{表2}
\end{center}
\end{table}

In the ABX preference test, participants compared the audio samples generated by the baseline model and the model proposed in this paper based on auditory perception, selecting their preferred audio. The test results are shown in Tab.~\ref{1}. The proposed method achieved significantly higher average preference scores than the baseline model, indicating that the generated audio has a higher quality in subjective auditory perception. Statistical results show that the p-values are below 0.05, validating the effectiveness of the model improvements. The MOS evaluation presents the results in Tab.~\ref{2}. The proposed model achieved a MOS score of \( 4.11\pm 0.12 \), surpassing the MOS of the baseline model. Notably, Sag-Tacotron is an AR TTS model, and the MOS results indicate that the speech generated by the proposed method is comparable to or even surpasses Sag-Tacotron in terms of naturalness. This demonstrates that the method proposed in this paper can generate highly natural speech within a NAR framework.

\begin{figure}[ht]
\centering
\includegraphics[width=0.5\textwidth]{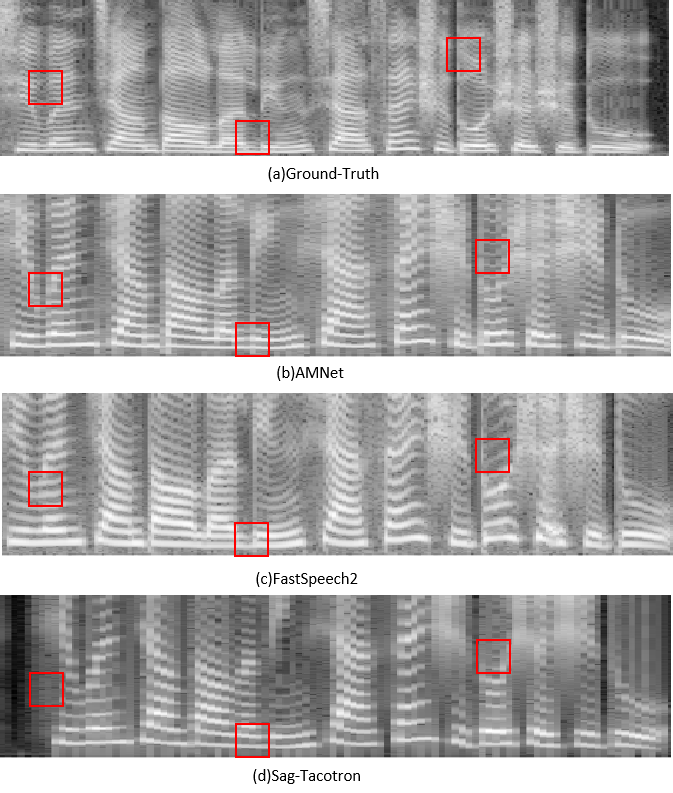}
\caption{Spectral details.}
\label{3}
\end{figure}

To more intuitively observe the differences between synthesized speech, we compared the features of the speech sample in the spectrogram, including pauses, stress, intonation, and others. As shown in Fig.~\ref{3}, we compared the spectrograms of natural recordings with those generated by the baseline models (FastSpeech 2 and Sag-Tacotron) and the model proposed in this paper. The results indicate that the speech generated by the proposed model is more consistent with natural speech in terms of pauses, stress, and intonation, making the synthesized speech sound more natural and realistic.

This paper highlights local details from the complete spectrogram, sampled from different time points. It focuses on pauses and details within the sentence, emphasizing their differences. From the comparison of the local information, there are significant differences in pause durations between Sag-Tacotron and FastSpeech 2. For instance, both Sag-Tacotron and FastSpeech 2 show noticeably prolonged pauses. However, the tone comparison in the far right column demonstrates that the proposed model performs better in tonal features than the baseline models. For example, in the audio, the tone is a stressed falling tone, whereas FastSpeech 2 generates a stressed rising tone. The tone generated by Sag-Tacotron is overly smooth and even, failing to exhibit natural tonal variations. In contrast, AMNet accurately displays the falling tone, which is consistent with natural intonation. These results indicate that the proposed model can more accurately infer audio pauses, durations, and tonal details.

\subsubsection{Objective evaluation}

\begin{table}[t]
\caption{Objective Evaluation Results}
\begin{center}
\begin{tabular}{|c|c|c|}
\hline
Model & MCD($\downarrow$) & $R^2(F0)(\uparrow)$ \\
\hline
Ground truth & / & / \\
\hline
FastSpeech2 & 6.28 & 0.784 \\
\hline
Sag-Tacotron & 5.78 & 0.607 \\
\hline
\textbf{AMNet (Ours)} & \textbf{5.54} & \textbf{0.874} \\
\hline
\end{tabular}
\label{表3}
\end{center}
\end{table}

In the objective evaluation, this paper uses MCD and fundamental frequency fitting goodness \( F0 (R^2) \) to assess the synthesized speech's quality quantitatively. MCD is an essential indicator for measuring speech feature distortion, with smaller values indicating higher speech quality. The experimental results show that compared to FastSpeech 2, the MCD of the proposed model is reduced by approximately \( 11.8\%\), indicating a significant improvement in the quality of the synthesized speech.

In the fundamental frequency (F0) analysis, this paper uses the YIN algorithm\cite{b37} to analyze the F0 curve of the audio. It employs the goodness of fit \( F0 (R^2) \) to evaluate the F0 gaps in the generated speech. The \( (R^2) \) value ranges from \( (-\infty ,1) \), with higher values indicating better F0 fitting performance. The experimental results show that the \( (R^2) \) value of the proposed model has improved compared to Sag-Tacotron, indicating that the proposed method is better at capturing the fundamental frequency characteristics of speech.

\begin{figure}[ht]
\centering
\includegraphics[width=0.5\textwidth]{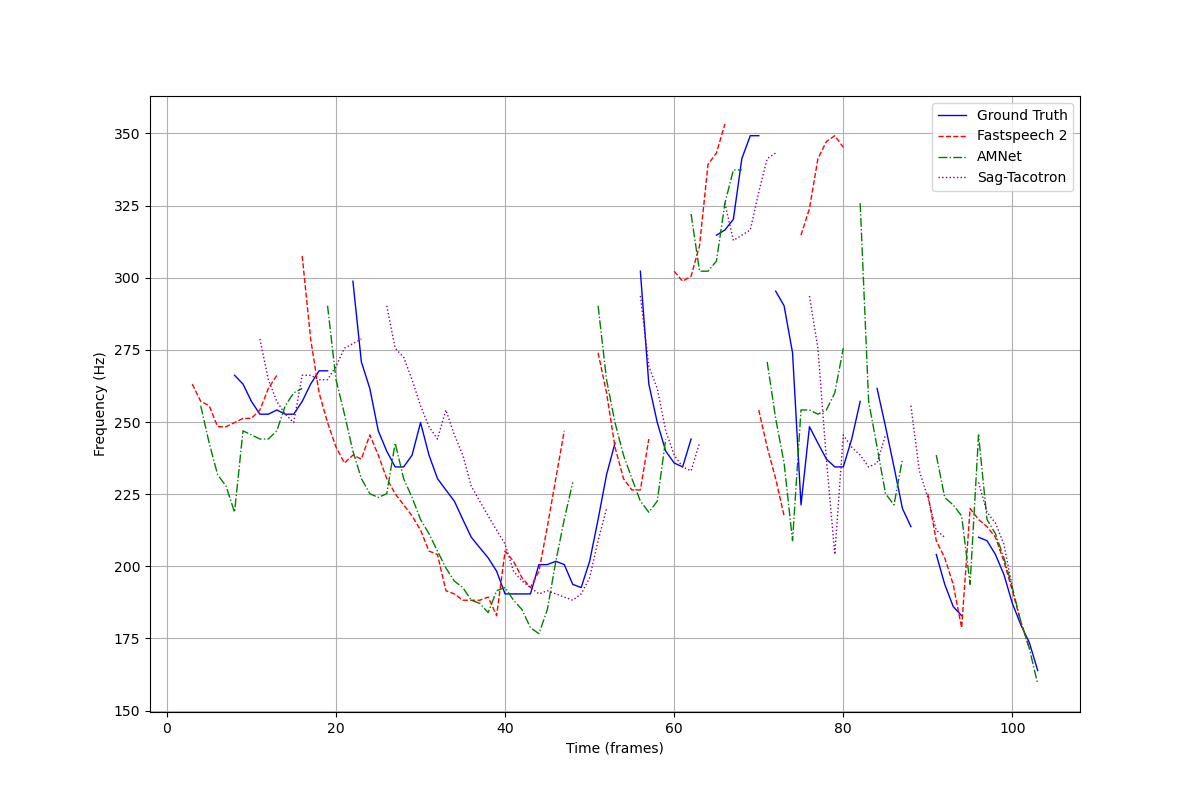}
\caption{The F0 curve comparison.}
\label{4}
\end{figure}

As shown in Fig.~\ref{4}, we plot the F0 contours of the actual audio, FastSpeech 2, Sag-Tacotron, and the audio generated by the model proposed in this paper. From the F0 curves, it can be observed that the F0 contour of FastSpeech 2 shows a noticeable shift in the latter half, consistent with the spectrogram results. Additionally, the F0 curve of Sag-Tacotron has a significantly wider fluctuation range than the other models, making the generated speech sound more monotone and resulting in the lowest \( (R^2) \) value. In contrast, the F0 curve of the proposed model outperforms the baseline models in terms of fluctuation range and fit, showing better alignment with the actual values. This also explains why the proposed model achieved higher MOS scores in the subjective evaluation.

Based on subjective and objective evaluation results, the model proposed in this paper significantly outperforms the baseline models across multiple metrics. The subjective evaluation results indicate that the speech generated by the proposed model exhibits higher naturalness and auditory quality, particularly regarding pauses, stress, and intonation details, which are closer to actual speech. The objective evaluation results show that the proposed model achieves better MCD and \( F0 (R^2) \)  values than the baseline models, validating the model's effectiveness in speech feature extraction and fundamental frequency modeling. These results demonstrate that the proposed method offers significant advantages in improving the performance of non-autoregressive speech synthesis tasks, providing valuable insights for future research.

\subsection{Ablation studies}
\begin{table}[t]
\caption{Ablation Experiment Results}
\begin{center}
\begin{tabular}{|c|c|c|c|}
\hline
\textbf{Model} & \textbf{MOS($\uparrow$)} \\
\hline
w/o LC (Local Convolution) & 3.96$\pm$0.12 \\
\hline
w/o Tone & 4.02$\pm$0.11 \\
\hline
w/o PDC & 3.97$\pm$0.12 \\
\hline
w/o LN & 3.95$\pm$0.12 \\
\hline
w/o PSP & 3.92$\pm$0.13 \\
\hline
\textbf{AMNet (Ours)} & \textbf{4.11$\pm$0.12} \\
\hline
\end{tabular}
\label{表4}
\end{center}
\end{table}

To analyze the specific impact of phrase structure annotation on model performance, we conducted a comparative experiment using Part-of-Speech Tagging (POST), a method widely applied in natural language processing. The experimental results are shown in Tab.~\ref{4}. The results indicate that phrase structure annotation plays a key role in model performance. In the experiment, we compared the impact of using PSP and Local Convolution (LC) on model scoring. When the model did not use PSP, the MOS was the lowest among all experiments, indicating phrase structure annotation's critical role in enhancing speech synthesis models' performance.
On the other hand, replacing PSP with POST also yielded good MOS scores, but overall performance was slightly inferior to that of PSP. This may be because the additional phrase labels from PSP allow for more accurate phrase boundary segmentation, thereby enhancing the model's ability to model contextual information. Furthermore, the phrase boundary information provided by PSP, combined with Multi-Scale Convolutional Channels, allows the model to be more sensitive to local contextual information, thus achieving better performance in speech synthesis tasks.

Ablation experiments under different settings further validate the effectiveness of the proposed technique. For example, the MOS score significantly decreases when layer normalization (LN) is not used. This indicates that LN effectively normalizes the scale of the input sequence and its position encoding, thereby improving model performance. Additionally, the experimental results demonstrate that introducing the Tone component contributes positively to the MOS score. Removing the Tone component leads to a decrease in the MOS score, reflecting the positive role of the Tone component in prosody and intonation modeling for speech synthesis tasks.

\section{Conclusions}
AMNet, the Acoustic Model Network proposed in this paper, offers a novel approach to improving Mandarin speech synthesis by integrating phrase structure annotation and local convolution mechanisms. Including tone decoupling further enhances the accuracy of pronunciation and tonal features. The experimental results show that AMNet outperforms traditional baseline models, achieving better subjective quality and objective performance metrics. This indicates that AMNet's design, which focuses on local context modeling and tone accuracy, significantly improves speech synthesis results. These findings highlight the potential of AMNet for advancing speech synthesis technology, providing a solid foundation for future research and development in the field of high-quality Mandarin speech synthesis.

\section*{Acknowledgements}

This study was funded by the Excellence Program Project of Tianshan, Xinjiang Uygur Autonomous Region, China (grant number 2022TSYCLJ0036), the Central Government Guides Local Science and Technology Development Fund Projects (grant number ZYYD2022C19), and the National Natural Science Foundation of China (grant numbers 62463029, 62472368 and 62303259).

\end{document}